# Birefringent anti-resonant hollow-core fiber

STEPHANOS YEROLATSITIS[1,2*], RILEY SHURVINTON[2], PENG SONG[3], YAPING ZHANG[3], ROBERT J. A. FRANCIS-JONES[1,4], KRISTINA R. RUSIMOVA[1,2*]

[1]*Centre for Photonics and Photonic Materials, University of Bath, Bath BA2 7AY, UK*
[2]*Department of Physics, University of Bath, Bath, BA2 7AY, UK*
[3]*School of Physics and Technology, University of Jinan, Jinan, 250022, Shandong, China*
[4]*Now at Clarendon Laboratory, University of Oxford, Parks Road, Oxford, OX1 3PU, UK*

*Corresponding author: s.yerolatsitis@bath.ac.uk; k.r.rusimova@bath.ac.uk

**Hollow-core fibers have demonstrated record performance in applications such as high-power pulse delivery, quantum computing, and sensing. However, their routine use is yet to become reality. A major obstacle is the ability to maintain the polarization state of light over a broad range of wavelengths, while also ensuring low attenuation and single-mode guidance. Here we simulated, fabricated and characterized a single-mode polarization-maintaining anti-resonant hollow-core fiber. The birefringence was achieved by introducing resonators of different thicknesses, thereby creating reduced symmetry in the structure. The measured group birefringence of $4.4 \times 10^{-5}$ at 1550 nm is in good agreement with the calculated group birefringence from the simulations. This corresponds to a phase birefringence of $2.5 \times 10^{-5}$ at 1550 nm. The measured loss of the fiber was 0.46 dB/m at 1550 nm. With its simple structure, low loss, and broadband operation this polarization-maintaining anti-resonant hollow-core fiber is a serious contender for applications in gas-based nonlinear optics and communications.**

## 1. INTRODUCTION

Air-guidance in hollow-core fibers holds the key to many exciting possibilities, e.g. the lowest loss signal transmission, [1,2] high-energy pulse delivery [3], deep-UV [4] and mid-IR [5] guidance, gas spectroscopy and gas lasers [6,7], and precise interferometric sensing [8]. However, their widespread practical use in many applications is hampered by their poor modal control, which in turn makes it fundamentally difficult to maintain the polarization state of the guided light. In solid core fibers, polarization control is achieved either by introducing stress rods in the cladding structure, or by forming shape anisotropy (e.g. by increasing the ellipticity of the core) [9]. Nevertheless, due to the inherent material properties of traditional silica fibers, their use has its limitations – namely, low damage threshold, material dispersion, and absorption. Hollow-core fibers offer a practical route to overcoming these limitations. However, in hollow-core fibers, light is confined within an air-core and therefore there is very little interaction with the dispersive silica cladding material. As a result, it is difficult to induce modal birefringence in these fibers with the methods applied to traditional solid-core fibers. Previously, high birefringence was demonstrated in hollow-core photonic bandgap fibers (HC-PBGF) by fabricating an elliptical core [10] or by modifying the cladding structure of the fiber to introduce polarization dependent loss [11]. Although HC-PBGFs have solved some of the issues associated with standard step-index fibers, they are complicated to fabricate, and their narrow operating bandwidth is a limiting factor.

Recent resurgent interest in hollow-core fibers has resulted in the development of anti-resonant hollow-core fibers [12]. The guiding mechanism in anti-resonant fibers is much simpler compared to that in photonic-bandgap fibers. The cladding structure of anti-resonant fibers comprises a ring of capillary resonators surrounding an air core. The walls of these capillaries act as Fabry-Perot resonators, confining the light in the core at their anti-resonant wavelengths. Numerical simulations of various anti-resonant structures have revealed the possibility of achieving shape birefringence in such fibers [13–16], but their practical demonstration remains elusive. Here, we report the first fabrication and characterization of a polarization-maintaining anti-resonant hollow-core fiber. The birefringence is achieved by modifying the thickness of two of six non-touching resonators (opposite to each other) that encompass the core, thereby reducing the rotational symmetry in the structure below threefold.

## 2. BREFRINGENT FIBER DESIGN

Standard, anti-resonant hollow-core fibers consist of a ring of capillary resonators around a hollow core, with all resonators of the same diameter ($d_1$) and thickness ($t_1$) (Design A). In our birefringent hollow-core fiber, the modal birefringence is achieved by changing the thickness, $t_2$ of two resonators (Design B).

Figure 1 shows a scanning electron microscope image of the hollow-core fiber fabricated based on Design B using the stack-and-draw method. All six resonator capillaries have a similar outer dimeter (~16 μm). A small variation in the diameter between the two sets of resonators of different thicknesses is due to the differential pressurization technique used to fabricate the fiber. Two of the resonator capillaries (opposite to each other, across the axis of the fiber) are three times thinner ($t_1$ = 210 nm) compared to the other four ($t_2$ = 610 nm). According to the antiresonant reflecting optical waveguide (ARROW) model, the four thicker capillary resonators will have their first anti-resonant guiding band centered at around 2400 nm, whereas the two thinner resonators will have their first anti-resonant band centered at around 800 nm [12]. The outer diameter of the fiber is 125 µm and the core diameter is 26 µm. In order to increase the interaction of the guided mode with the surrounding glass structure, the core size was intentionally selected such that it is half the size that typically ensures minimal attenuation at 1550 nm [17]. The resonator diameter to core diameter ratio was selected to ensure single mode operation of the fiber [18].

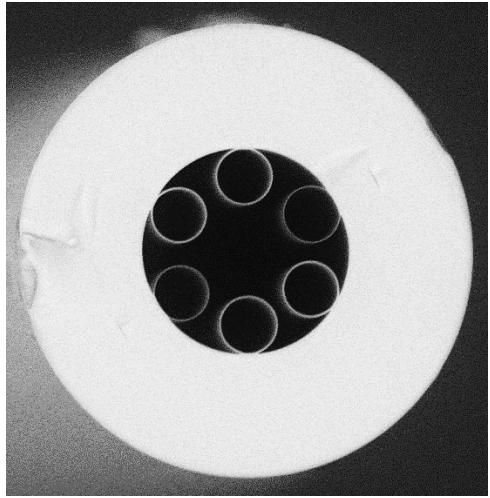

Fig. 1. Cross-sectional SEM (scanning electron microscope) image of the birefringent hollow-core fiber. The outer diameter of the fiber is 125 µm and the core diameter is 26 µm.

|  | Design A | Design B |
| --- | --- | --- |
| Core diameter (µm) | 24 | 26 |
| Thickness, $t_1$ (nm) | 610 | 610 |
| Thickness, $t_2$ (nm) | - | 210 |
| Resonator diameter, $d_1$ (µm) | 15.6 | 16.7 |
| Resonator diameter, $d_2$ (µm) | - | 16.1 |

Table 1: Parameters of the anti-resonant hollow-core fibers.

Table 1 summarizes the parameters of fiber Design B as measured from the SEM image, and introduces a standard anti-resonant fiber (Design A) of the same outer diameter and first anti-resonant band centered at 2400 nm. Using the parameters in Table 1 and a finite element method solver (Comsol Mutiphysics 5.3a), we simulated both fiber designs and obtained the effective refractive index of the fundamental mode in the two orthogonal polarizations ($n_x$ and $n_y$) across a range of wavelengths. Fig. 2 shows a schematic diagram of each of the fiber designs from Table 1 (a-b), together with their corresponding confinement loss (c-d) and phase birefringence $B_P = (|n_y - n_x|)$ (e) as a function of wavelength. As per the design, both fibers have an anti-resonant guiding band centered at ~ 800 nm and minimum attenuation of ~ 0.01 dB/m at 780 nm. For both fibers, there is also another anti-resonant guidance band at longer wavelengths, above 1300 nm. The minimum confinement loss in this band for Design A is 0.95 dB/m at 1740 nm, compared to ~ 0.7 dB/m at 1550 nm for Design B. It is important to note the difference in the width of this transmission band between the two designs. We can attribute the much narrower transmission band of Design B at longer wavelengths to the two different resonator thicknesses used in this design, with the thinner resonators completely lacking a transmission band at longer wavelengths.

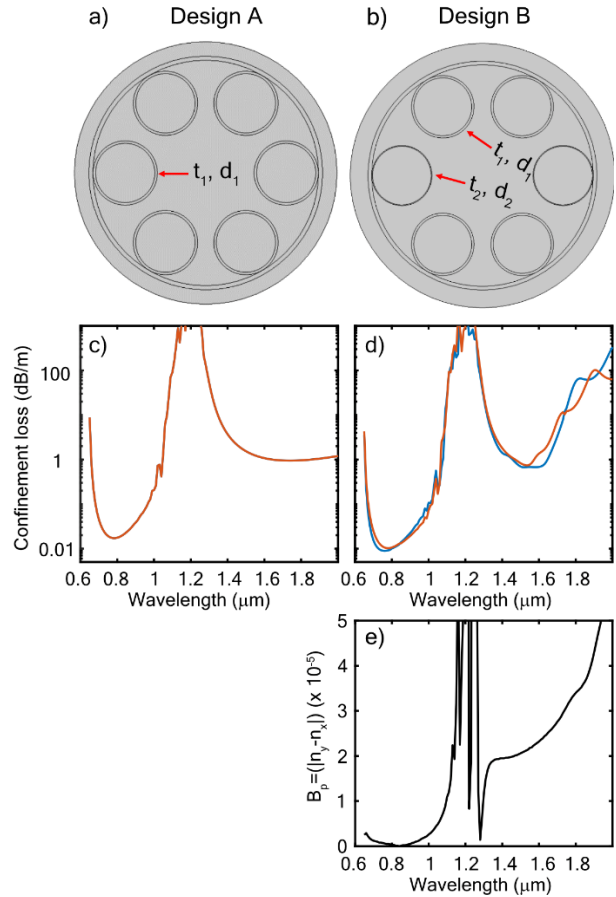

Fig. 2 Schematic diagram of the simulated fiber based on (a) Design A, and (b) Design B. Calculated confinement loss (for the two orthogonal polarizations) of (c) Design A, and (d) Design B. (e) Calculated phase birefringence as a function of wavelength for Design B.

In Design A, as expected, the calculated phase birefringence of this standard fiber design is zero and both polarizations have identical confinement losses as a function of wavelength. By contrast, in Design B, there is a difference between the effective refractive indices of the two polarization modes resulting in the calculated phase birefringence shown in Fig. 2. (e). There is a clear difference in the calculated phase birefringence in each transmission band with $B_P \sim 10^{-5}$ at 1550 nm – significantly higher than the birefringence in the transmission band centered at 800 nm. In addition, there is a small difference in the calculated confinement loss for each polarization which is more noticeable towards the longer wavelength range.

## 3. RESULTS

The cutback method was used to determine the transmission bands and the attenuation of the fabricated fiber of Design B. White light from a Xe light source was coupled into the core of 50 m of the fiber and the transmission was recorded using an OSA. The fiber was cut back to 10 m and the transmission was recorded again. Fig 3 shows the resulting attenuation spectrum of the fiber. We measured attenuation of 0.46 dB/m at 1550 nm.

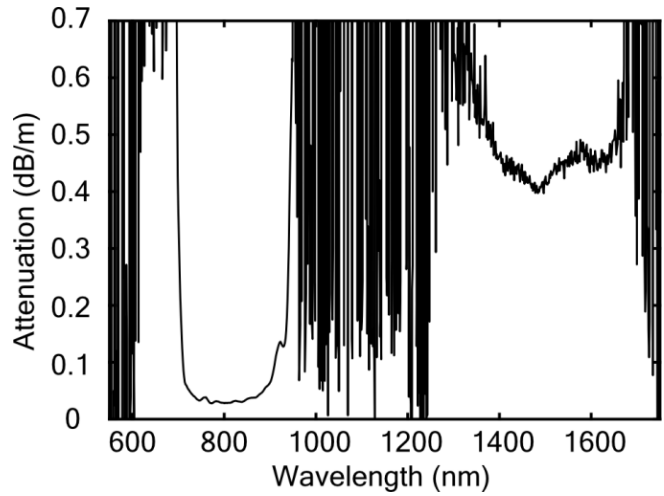

Fig. 3. Spectral attenuation of the birefringent hollow-core fiber.

Due to the very large anticipated beat length, it is not trivial to measure the phase birefringence of our hollow-core fiber. Instead, we measured the group birefringence $B_G = B_P - \lambda\, dB/d\lambda$ of our fibre using the group beatlength technique [19]. For the beatlength measurement, light from a broadband light source (near-IR edge-emitting LED source Agilent 83487a of spectral range 1100 nm -1600 nm) was coupled into a 10 m length of the hollow-core fiber, as shown in Fig. 4(a). As the light from the light source is unpolarized, a polarizer was introduced at the output of the light source. A broadband quarter-wave plate was also used to ensure that both polarization states of the hollow-core fiber are excited. At the output, a polarizer and an optical spectrum analyzer (OSA) were used to analyze the light from the hollow-core fiber. The output polarizer was set to ensure the strongest beat pattern resolved at the OSA, corresponding to 45° rotation of the polarizer with respect to the birefringent axis of the hollow-core fiber. An NIR camera was also used to capture the near-field intensity pattern of the output of the fiber. The near-field intensity distribution (Fig. 4(b)) resembled a pure fundamental mode, confirming the single mode performance of the fiber.

Due to the birefringence of the fiber, the output light from the hollow-core fiber shows a strong characteristic spectral fringe pattern (Fig. 5(a)) when observed with the OSA. For a fibre of length $L$, the group birefringence ($B_G$) can be calculated using the spectral spacing of the fringes ($\Delta\lambda$) and equation:

$$B_G = \frac{\lambda_0^2}{L\Delta\lambda}, \qquad (1)$$

with $\lambda_o$ the central wavelength [6]. The fiber was cut down by two meters and the measurement repeated. As expected, the fringe spectral spacing increased.

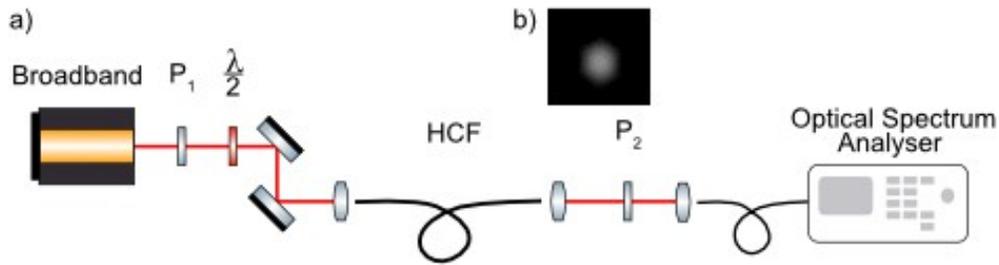

Fig. 4. (a) Experimental setup for measuring the fringe beat pattern of the two polarization states of the fiber. (b) Near-field intensity pattern at the output of the fiber at 1550 nm.

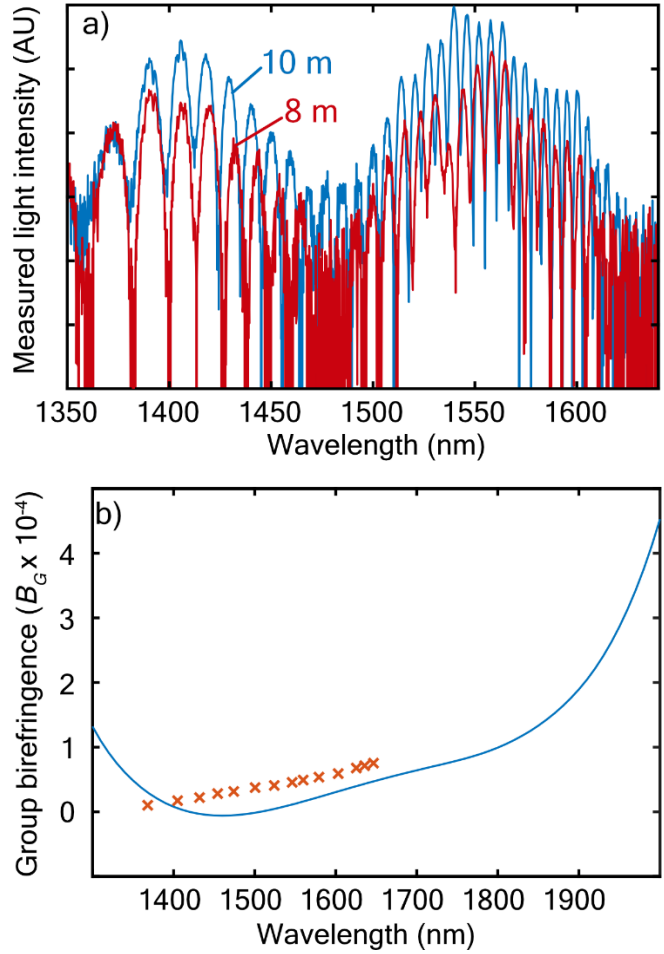

Fig. 5. (a) Spectral fringe pattern using a 10 m (in blue) and an 8 m (in red) piece of the birefringent hollow-core fibre. (b) Simulated (blue) and experimental (orange) group birefringence as a function of wavelength.

Using the procedure in Ref. [10] and the data for the simulated Design B (by fitting a polynomial), we obtained the group birefringence (Fig. 5(b), blue) for the range of wavelengths that we are interested in (between 1300 nm and 2000 nm). Fig. 5(b) also shows the experimental values of $B_G$ calculated from the beatlength measurement using Equation (1). The experimental and the simulated data follow the same trend, with a small offset in wavelength. This offset can be attributed to the accuracy of the resonator thicknesses that were measured from the SEM image in Fig. 1 and then used for the simulation (Table 1). Nevertheless, there is a clear correlation between the experimental and the simulated data. The measured magnitude of the group birefringence for the hollow-core fiber is $B_G = 4.4 \times 10^{-5}$ at 1550 nm, and the simulated value of the phase birefringence is around $B_P = 2.5 \times 10^{-5}$ at the same wavelength.

## 4. DISCUSSION AND CONCLUSSION

The calculated phase birefringence of our polarization-maintaining anti-resonant hollow-core fiber is two orders of magnitude lower compared to that of highly birefringent HC-PBGFs [19], and one order of magnitude lower compared to the phase birefringence of commercially available solid core polarization-maintaining fibers [9]. However, in HC-PBGFs the high birefringence is traded off for transmission losses that are three times higher than those of our fiber. This high loss coupled with a very narrow transmission band, render HC-PBGFs a poor choice especially for broadband nonlinear processes [20]. On the other hand, the glass core of conventional solid-core polarization maintaining fibers restricts their use in applications requiring pulsed light. Therefore, with its simple design, broadband operation, and air-guidance a polarization-maintaining anti-resonant hollow-core fiber is a significant leap forward.

To our knowledge, this is the first demonstration of a birefringent anti-resonant hollow-core fiber. By using resonators of two different thicknesses we achieved a phase birefringence of $\sim 10^{-5}$ at 1550 nm. Each set of resonators have their first guidance band at around 800 nm and at around 2400 nm, respectively. We believe that the measured polarization-maintaining properties of this fiber at 1550 nm arise from the enhanced interaction of the guided light with the surrounding structure. At 1550 nm, the fiber operates at the lower edge of the transmission band centered at 2400 nm for the thicker resonators, and approaches the resonance at ~1200 nm. Light guided at 1550 nm is also at the upper edge of the transmission band centered at 800 nm for the thinner resonators, where losses due to the small core size dominate [17]. Moreover, maintaining the loss of the fiber below 1 dB/m is crucial for the potential use of the fiber in various applications, such as polarization-maintaining fiber amplifiers [21] and sensing [22]. Stronger birefringence may be achieved [13,14] by introducing a larger thickness or size dissimilarity between the resonators. Lower loss may also be possible for larger core-sizes. However, the trade-off between attenuation loss and birefringence in hollow-core fibers is unavoidable.

**Funding.** We acknowledge support from the UK Engineering and Physical Sciences Research Council (EP/M013243/1, EP/K03197X/1 and EP/S001123/1) and Innovate UK project FEMTO-AAD (102671).

**Acknowledgment**. We would like to thank Tim Birks, William Wadsworth, David Bird, James Stone and Peter Mosley for the fruitful discussion